\documentclass[options]{JHEP3}
\usepackage{amsmath,amssymb}
\usepackage[dvips]{graphics}
\def\beq{\begin{equation}}
\def\eeq{\end{equation}}
\def\bea{\begin{eqnarray}}
\def\eea{\end{eqnarray}}
\title{On the AdS Higher Spin / O(N) Vector Model Correspondence: degeneracy of the
holographic image}
\author{Danilo E. Diaz and Harald Dorn
\\ Humboldt-Universit\"at zu Berlin, Institut f\"ur Physik
\\Newtonstr.15, D-12489 Berlin
\\E-mail: \email{ddiaz,dorn@physik.hu-berlin.de}}

\abstract{We explore the conjectured duality between the critical
O(N) vector model and minimal bosonic massless higher spin (HS)
theory in AdS. In the free boundary theory, the conformal partial
wave expansion (CPWE) of the four-point function of the scalar
singlet bilinear is reorganized to make it explicitly
crossing-symmetric and closed in the singlet sector, dual to the
bulk HS gauge fields. We are able to analytically establish the
factorized form of the fusion coefficients as well as the
two-point function coefficient of the HS currents. We insist in
directly computing the free correlators from bulk graphs with the
unconventional branch. The three-point function of the scalar
bilinear turns out to be an ``extremal'' one  at $d=3$. The
four-point bulk exchange graph can be precisely related to the
CPWs of the boundary dual scalar and its shadow. The flow in the
IR by Legendre transforming at leading 1/N, following the pattern
of double-trace deformations, and the assumption of degeneracy of
the hologram lead to the CPWE of the scalar four-point function at
IR. Here we confirm some previous results, obtained from more
involved computations of skeleton graphs, as well as extend some
of them from $d=3$ to generic dimension $2<d<4$.}

\preprint{HU-EP-06/10}

\begin{document}
%------------------------------------------------------------------------
\section{Introduction}
Duality between strongly coupled SYM in the large N limit and
weakly coupled SUGRA is one of the forms of Maldacena's
conjecture. Many interesting results have been obtained and many
tests have been performed in this regime. However, much less is
known about the bulk dual to perturbative gauge theories or even
to free theory. It has been conjectured that the dual bulk theory
to large N free gauge theory is a HS theory of Fradkin-Vasiliev
type~\cite{Wit01, SS02}. A simpler scenario for testing these
ideas has been proposed by Klebanov and Polyakov~\cite{KP02},
concerning the bulk dual of the critical O(N) vector model. Vector
models have always been useful in understanding features that
arise in the more complicated case of gauge theories. Here one can
use the vast experience in large-N limit of O(N) vector models to
reconstruct the bulk theory. An analogous attempt has been started
by Gopakumar~\cite{Gop03} for the singlet bilinear sector
(twist-two operators in $d =4$) of the gauge theory; but despite
the initial success in casting two and three-point function of
scalar bilinears into AdS amplitudes, four-point correlators have
remained a challenge~\footnote{See however \cite{AKR06} for recent
progress and for a large set of references to proposals for string
dual of large N gauge theories.}. The technical difficulty has
been that one should include the whole tower of HS fields, dual to
the HS conserved currents of the CFT at the boundary, in the
exchange graphs since the OPE structure of the free field
correlators involve the whole tower of conserved currents. A bulk
theory consistently truncated to massless fields should be
reflected somehow in a closure of the corresponding dual sector of
CFT operators. We study the free four-point function of the scalar
bilinear by means of a conformal partial wave expansion and
reorganize it so as to involve only the minimal twist
sector~\footnote{Here we follow a suggestion in~\cite{SS02} for
the free gauge theory case and turn it into a quantitative
result.}, i.e. the conserved HS currents and their descendants, as
required by the correspondence. The fusion coefficients are
analytically checked to factorize in terms of two- and three-point
function coefficients. A comparison with the correlators at the IR
fixed point can be made by means of the amputation procedure that
realizes the Legendre transformation connecting the two conformal
theories at leading order in the large-N expansion, in close
analogy with the effect of double-trace deformations in the gauge
theory ( see e.g.~\cite{GK03} and references therein). We pursue
the view that both CFTs are on equal footing, related by Legendre
transformation, and that one can compute directly the Witten
graphs with either branch $\Delta_{+/-}$ ~\footnote{While
preparing this text a paper by Hartman and Rastelli~\cite{HR06}
appeared which also stresses this view.}. Our aim is to have an
autonomous way to compute directly in the free theory, having in
mind a possible extension to free gauge theories, with no need of
Legendre transforming from a conjugate CFT that arises at leading
large N but whose existence is otherwise uncertain. As a
consequence, at $d=3$ one has a vanishing three-point function for
the scalar bilinear at IR. On the other hand, at UV the
three-point function is nonzero due to the compensation of the
vanishing coupling by a divergence of the corresponding Witten
graph. This is similar to the case of extremal correlators, see
e.g. \cite{D'HF02}. The underlying assumption of a common bulk
theory, degeneracy of the holographic image, is also consistent
with the CPWE of the four-point correlators. Progress in the bulk
side of the correspondence is considerably more difficult due to
the complicated nature of the interacting HS theories on AdS. We
use the CPWs to mimic the effect of the corresponding bulk
exchange graphs, even though the CPW is generically only a part of
the Witten graph and one can only hope that after including the
whole tower of HS exchange the additional terms cancel out. In
this direction, we study the scalar exchange in AdS and relate it
to the corresponding CPW.

The paper is organized as follows: we start by briefly describing the
conjecture. Then we define the higher spin conserved currents and
compute some of their correlators~\footnote{In particular, the
coefficient of the two-point function is analytically obtained. In
the literature, this result is only known as extrapolation of
 an expression obtained by computer algebraic
 manipulations~\cite{Ans99}.}.
We then obtain the conformal partial waves of the HS currents to
study
 the scalar bilinear four-point function, in an attempt to get closer to a bulk AdS
 formulation, and manipulate the results to achieve the closure of
 the bilinear sector. We then move to the IR critical theory by
 means of the amputation procedure and, based on the holographic
 degeneracy, predict the CPWE at IR including fusion coefficients.
 Finally, we explore at $d=3$ the precise connection between the
 scalar exchange Witten graph and the corresponding scalar CPW of
 dimension $\Delta_-=1$. Some useful formulas and details of the calculations
 are collected in various appendices.
%---------------------------------------------------------------------------
\section{The Klebanov-Polyakov Conjecture}

Let us briefly review the essentials of the conjecture. The
singlet sector of the critical 3-dim O(N) vector model with the
$(\overrightarrow{\varphi}^2)^2$ interaction is conjectured to be
dual, in the large N limit, to the minimal bosonic theory in
$AdS_4$ containing massless gauge fields of even spin. There is a
one-to-one correspondence between the spectrum of currents and
that of massless higher-spin fields. In addition we have a scalar
bilinear J mapped to a bulk scalar $\phi$. The AdS/CFT
correspondence working in the standard way (conventional dimension
$\Delta_+$ for J) produces the correlation functions of the
singlet currents in the interacting large N vector model at its IR
critical point from the bulk action in $AdS_4$ by identifying the
boundary term $\phi_0$ of the field $\phi$ with a source in the
dual field theory (cf. appendix F). At the same time, the
correlators in the free theory are obtained by Legendre
transforming the generating functional with respect to the source
that couples to the scalar bilinear J; this corresponds on the AdS
side to the procedure for extracting the correlation functions
working with the unconventional branch
$\Delta_-$~\cite{KP02,KW99}. However, we want to stress that one
can directly compute the bulk graphs with the $\Delta_-$ branch,
by using the boundary term $A$ (cf. appendix F) as source in the
boundary theory, and that the boundary correlator obtained is
precisely related by Legendre transformation to the one computed
with the standard $\Delta_+$ branch.

%---------------------------------------------------------------------------
\section{Free O(N) Vector Model}

We start by considering N elementary real fields $\varphi^a$  in
d-dimensional Minkowski space, vectors under the global O(N)
symmetry and Lorentz scalars with canonical scaling dimension
$\delta=d/2-1$(in what follows we switch to Euclidean space). They
satisfy the free equation of motion $\partial^2 \varphi^a=0$. We
normalize their two-point function as \beq \langle
\varphi^a(x_1)\varphi^b(x_2)\rangle=\frac{\delta^{ab}}{r_{12}^{\delta}}\,,
\quad a,b=1,...,N\;, \eeq where $\quad
r_{ij}=|x_i-x_j|^2=|x_{ij}|^2$.

%----------------------------------------
\subsection{HS Conserved Currents}

In this free theory there is an infinite tower of (even) higher
spin currents, bilinear in the elementary fields, which are
totally symmetric, traceless and conserved. These three properties
fix their form, their precise expression can be found
in~\cite{Ans99,LMR04}. We will only need them in the following
form (assuming normal order and omitting free indices)
\beq
J_l=\sum^{l}_{k=0}a_k\;\partial^k\overrightarrow{\varphi}\cdot
\partial^{l-k}\overrightarrow{\varphi}\;-traces,
\eeq with~\footnote{The Pochhammer symbol
$(q)_r=\frac{\Gamma(q+r)}{\Gamma(q)}$. } \beq
a_k=a_{l-k}=\frac{1}{2}(-1)^k {l \choose k} \frac{(\delta)_l}
{(\delta)_k(\delta)_{l-k}}. \eeq
Note that this convention means
\beq J_l=\overrightarrow{\varphi}\cdot\partial^{l}
\overrightarrow{\varphi}\;+.., \eeq
where the ellipsis stands for
terms involving derivatives of both fields. They are conformal
quasi-primaries, minimal twist operators with scaling dimension
\beq \Delta_l=d-2+l=2\delta+l. \eeq The AdS/CFT Correspondence
relates them to massless HS bulk field since the canonical
dimension $\Delta_l$ precisely saturates the unitarity bound for
totally symmetric traceless rank $l$ tensors ($l>0$, even).

%-----------------------------------
\subsection{Two- and Three-Point Functions}

The singlet bilinear sector is completed by adding to the above
list the scalar bilinear $J=\overrightarrow{\varphi}^2$,
spin-zero current, with canonical dimension
$\Delta_J=d-2=2\delta$. At $d=3$ its bulk partner is a conformally
coupled scalar.

Let us compute the two-point function of the HS currents and the
three-point function of two spin-zero and a HS current.

The conformal symmetry fixes the form of the two-point function up
to a constant~(\ref{two}), \beq \langle J_{l\;{\mu_1...\mu_l}}(x)
J_{l\;{\nu_1...\nu_l}}(0)\rangle = C_{J_l}\;r^{-2\delta-l}
\;sym\{I_{\mu_1\nu_1}(x)...I_{\mu_l\nu_l}(x)\}. \eeq To find the
coefficient it is then sufficient to look at the term $2^l
\frac{x...x}{r^l}$ involving $x$ $2l$ times. By Wick contracting
we get \beq \langle J_{l}(x)
J_{l}(y)\rangle=\sum^l_{k,s=0}a_k\,a_s\{\partial^k_x\partial^s_y
\langle \varphi^a(x)
\varphi^b(y)\rangle\partial^{l-k}_x\partial^{l-s}_y \langle
\varphi^a(x) \varphi^b(y)\rangle + (s\leftrightarrow
l-s)\}-\mbox{traces}. \eeq Using now the symmetry $a_k=a_{l-k}$,
trading $\partial_y$ by $-\partial_x$ and taking $y=0$ we get, up
to trace terms, \beq 2N\sum^l_{k,s=0}a_k\,a_s\;(\partial^{k+s}
r^{-\delta})\;(\partial^{2l-k-s}
r^{-\delta})=2^{2l+1}N\frac{x..x}{r^{2\delta+2l}}\sum^l_{k,s=0}a_k\,a_s\;
(\delta)_{k+s}\,(\delta)_{2l-k-s}. \eeq The double summation is
done using generalized hypergeometric series in appendix~\ref{HS2}
giving $\frac{1}{4}l! (2\delta-1+l)_l$. The coefficient of the
two-point function is then ($l>0$) \beq \label{C2}
C_{J_l}=2^{l-1}Nl!(2\delta-1+l)_l\, \eeq that coincides with the
extrapolated result reported by Anselmi~\cite{Ans99}. Analogously,
the three-point function form is dictated by the conformal
symmetry~( see \ref{three1}) \beq \langle J(x_1) J(x_2)
J_{l\;{\nu_1...\nu_l}}(x_3)\rangle = C_{JJJ_l}\;(r_{12}\,r_{23}\,r_{31})^{-\delta}
\;\lambda^{x_3}_{\mu_1..\mu_l}(x_1,x_2). \eeq

Now we focus on the coefficient of $2^l \frac{x_{31}...x_{31}}{r^l}$
involving $x_{31}$ $2l$ times after Wick-contracting,
\[
\langle J(x_1) J(x_2) J_l(x_3)\rangle=\langle \varphi^a(x_1)
\varphi^b(x_2)\rangle \langle \varphi^b(x_2)
\varphi^c(x_3)\rangle\partial^l_{x_3}\langle \varphi^c(x_3)
\varphi^a(x_1)\rangle+permutations +...\] \beq =4N
(r_{12}r_{23})^{-\delta}2^l(\delta)_l\frac{x_{31}...x_{31}}
{r^{\delta+l}_{31}}+...\;. \eeq Finally, we get for the
coefficient of the three-point function ($l>0$)
\beq \label{C3}
C_{JJJ_l}=2^{l+2}N(\delta)_l\;. \eeq

The corresponding values for the scalar are $C_J=2N$ and $C_{JJJ}=8N$.
%-----------------------------------
\subsection{Scalar Four-Point Function}
\begin{figure}
\begin{center}
\resizebox{110mm}{!}{\includegraphics{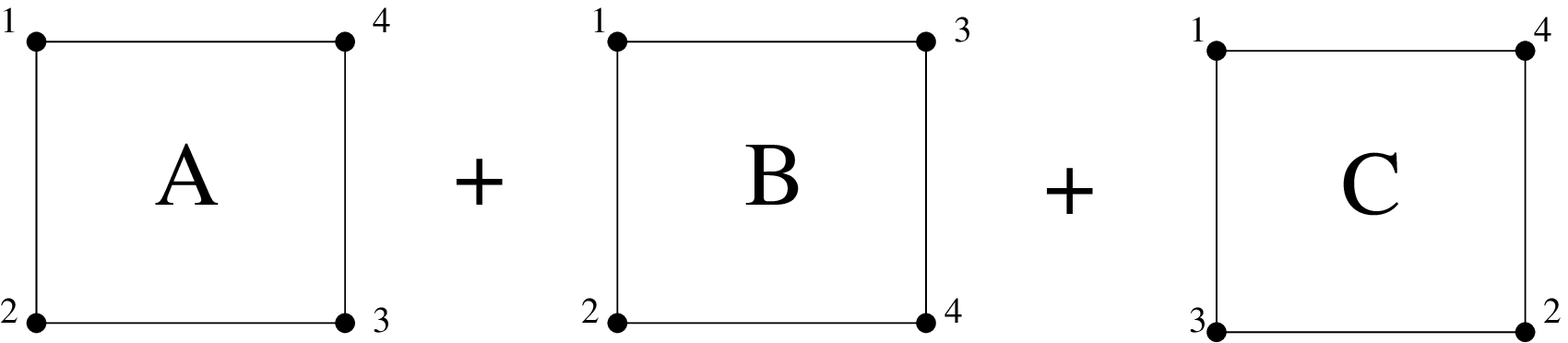}}
\end{center}
\caption{Connected part of the free four-point function of the
spin-zero current $J$.}\label{f1}
\end{figure}

The four-point function  contains much more dynamical information
encoded in a function of two conformal invariant cross-ratios
which is not fixed by conformal symmetry. Still, its form is
constrained by the OPE of any two fields and therefore the
contributions of operators of arbitrary spin, including their
derivative descendants, can be unveiled. The connected part of the
spin-zero current four-point function is obtained by Wick
contractions \beq \label{four-free}\langle
J(x_1)J(x_2)J(x_3)J(x_4)\rangle_{free, conn}
=\frac{16N}{(r_{12}r_{34})^{2\delta}}\left\{
u^{\delta}+(\frac{u}{v})^{\delta} +
u^{\delta}(\frac{u}{v})^{\delta}\right\}, \eeq where
$u=\frac{r_{12}\,r_{34}}{r_{13} \,r_{24}}$ and $
v=\frac{r_{14}\,r_{23}}{r_{13}\,r_{24}}$. Diagrammatically, it is
given by the three boxes in fig.~\ref{f1}.

Following Klebanov and Polyakov \cite{KP02} we notice that the
leading term in the box diagram
\beq
\frac{1}{(r_{12}r_{23}r_{34}r_{41})^{1/2}}\sim
\frac{1}{(r_{12}r_{34})^{1/2}}\frac{1}{r_{13}} \eeq
in the direct
channel limit $1\rightarrow2, 3\rightarrow4 $ correctly reproduces
the contribution of the scalar $J$ with dimension
$\Delta=\Delta_J=1$ to the double OPE (see, e.g. \cite{D'HF02}),
which in general reads
\beq \langle
J(x_1)J(x_2)J(x_3)J(x_4)\rangle
\sim\frac{1}{(r_{12}r_{34})^{\Delta_J-\Delta/2}}\frac{1}{(r_{13})^{\Delta}}.
\eeq

Sub-leading terms in the expansion of the box diagram should
correspond to the contribution of the currents
$J_l\sim\overrightarrow{\varphi}\cdot\partial^l\overrightarrow{\varphi},
l>0$. This structure is precisely what we want to study in detail
and the best way to identify all these contributions is via a
conformal partial wave expansion (CPWE); i.e., decomposing into
eigenfunctions of the quadratic Casimir of the conformal group
SO(1,d+1) in Euclidean space $\mathbb{R}^d$ \cite{DO03}.

%---------------------------------------------------------------------------
\section{From Free Fields to AdS via CPWE}

The attempts to cast the box diagrams into AdS amplitudes via
Schwinger parametrization have not succeeded so far~\cite{Gop03}.
From the previous analysis of the OPE, it becomes apparent that
the whole tower of HS field exchange has to be taken into account.
Even though some progress has been made in obtaining bulk to bulk
propagators for the HS fields in AdS~\cite{LMR03}, there is no
closed analytic form that could be used to include all the
infinite tower. We will content ourselves with the CPW amplitude
to mimic the effect of the corresponding exchange Witten graph. In
general, the CPW is contained in the exchange Witten graph but
there appear additional terms that cannot be precisely identified
as CPWs~\cite{LT98,Liu98,HPR00,HPR000,DO00}.

Let us first quote the essentials of the CPWE~(see, e.g.,
\cite{DO00} and references therein). The contribution of a
quasi-primary $O^{(l)}_{\mu_1...\mu_l}$ of scale dimension
$\Delta$ and spin $l$, and its derivative descendants, to the OPE
of two scalar operators $\phi_i$ of dimension $\Delta_i$,
\beq
\label{OPE} \phi_1(x)\phi_2(y)\sim \frac{C_{\phi_1\phi_2
O^{(l)}}}{C_{O^{(l)}}}\frac{1}{\mid x-y\mid^{\Delta_1+\Delta_2
-\Delta+l}}C^{(l)}(x-y,\partial_y)_{\mu_1..\mu_l}O^{(l)}
_{\mu_1...\mu_l}(y). \eeq
The derivative operator is fixed by
requiring consistency of the OPE with the two- and three-point
functions of the involved fields (see Appendix A). Based on these
constraints one can work out the contribution of the conformal
block corresponding to the quasi-primary $O^{(l)}$, and its
derivative descendants, to the four-point function. This is given
by the Conformal Partial Wave (see appendix~\ref{CPW-rec})
\bea
\label{CPW-fusion}
\langle\phi_1(x_1)\phi_2(x_2)\phi_3(x_3)\phi_4(x_4) \rangle & \sim
& \frac{C_{\phi_1\phi_2 O^{(l)}}C_{\phi_3\phi_4
O^{(l)}}}{C_{O^{(l)}}}(\frac{r_{24}}{r_{14}})^{\Delta_{12}/2}
(\frac{r_{14}}{r_{13}})^{\Delta_{34}/2}
\frac{u^{(\Delta-l)/2}}{(r_{12}r_{34})^{\Delta_{\phi}}}\nonumber\\
&&\;\times
G^{(l)}(\frac{\Delta-\Delta_{12}-l}{2},\frac{\Delta+\Delta_{34}-l}
{2},\Delta;u,v), \eea
which depends on the two
cross-ratios
\beq u=\frac{r_{12}\,r_{34}}{r_{13}\,r_{24}}\quad ,\qquad
v=\frac{r_{14}\,r_{23}}{r_{13}\,r_{24}}. \eeq

The CPWs have been obtained as double series in the direct channel
limit $(u,1-v)\rightarrow 0$ by several authors (see, e.g.
~\cite{LR93} and references therein). They can also be shown to
satisfy the recurrence relation (\ref{recurrences}), obtained by
Dolan and Osborn~\cite{DO00}.

%-----------------------------------
\subsection{CPWs of the HS Free Currents}

We are interested in the singlet bilinears, minimal twist
operators, in the free O(N) vector model. These are the spin-zero
($J\sim \varphi^a\varphi^a$) and the higher spin conserved
currents ($J_l\sim \varphi^a\partial^l\varphi^a$) with canonical
dimension $\Delta_l=d-2+l$. We can consider this limiting case in
the recurrences~\ref{recurrences} by first setting
$e=(\Delta-l)/2=d/2-1=\delta$ and in the end $S=2\delta+l$,
$b=\delta$. A crucial simplification occurs in the recurrence
relation, only the third line in~(\ref{recurrences}) survives:
\beq G^{(l)}(b,\delta,S;u,v)
=\frac{1}{2}\frac{S+l-1}{\delta+l-1}\left\{
G^{(l-1)}(b,\delta,S;u,v)-G^{(l-1)}(b+1,\delta,S;u,v)\right\} \eeq
 The iteration can then be easily done for the coefficients of
 the double expansion
 \beq
G^{(l)}(b,e,S;u,v)=\sum^{\infty}_{m,n=0}a^{(l)}_{nm}(b,S)\,
\frac{u^n}{n!}\frac{(1-v)^m}{m!}\;. \eeq
Pascal's triangle
coefficients ${l\choose k}$ arise to get
\beq
a^{(l)}_{nm}(b,S)=\frac{1}{2^l}\frac{(S)_l}{(\delta)_l}
\sum^{l}_{k=0}(-1)^k{l \choose k}a^{(0)}_{nm}(b+k,S), \eeq
where
$\delta=\mu-1=d/2-1$ and we start with the scalar exchange
~\ref{scalar}
\beq
a^{(0)}_{nm}(b,S)=\frac{(\delta)_{m+n}}{(S)_{m+2n}}(S-b)_n(b)_{m+n}.
\eeq
This can be summed up into a closed form involving a
terminating generalized hypergeometric of unit argument
\beq
a^{(l)}_{nm}(b,S)=\frac{1}{2^l}\frac{(\delta+l)_{m+n-l}}{(S+l)_{m+2n-l}}
(b)_{m+n}(S-b)_n
 \;\,_3F_2{{-l,1+b-S,b+m+n}\choose {b,1+b-S-n}}.\eeq
With these conventions, the normalization is fixed by
\beq
a^{(l)}_{0l}=(-\frac{1}{2})^l l!. \eeq
Now, using twice an identity (\cite{AAR}, pp.141), obtained as a limiting case
of a result due to
Whipple for balanced $_4F_3$ series, one can rewrite the
coefficients as a terminating (after $n+1$ terms) series. This
coincides with the result from the ``Master Formula'' in
~\cite{LMR02}~\footnote{In their
  normalization, $a^{(l)}_{0l}$ are set to 1.} for $b=\delta$ and $S=2\delta+l$
\[
a^{(l)}_{nm}=a^{(l)}_{0l}{m+n\choose
l}\frac{(\delta+l)^2_{m+n-l}}{(2\delta+2l)_{m+n-l}}
 \;\,_3F_2{{-n,1+m+n,\delta+m+n}\choose {1+m+n-l,2\delta+m+n+l}}\]
 \beq
 \label{cpwcoeff}
 =a^{(l)}_{0l}\sum^{n}_{s=0}(-1)^s{n \choose s}{m+n+s\choose
l}\frac{(\delta+l)_{m+n-l}(\delta+l)_{m+n-l+s}}{(2\delta+2l)_{m+n-l+s}}.
\eeq
In this form one can easily recognize a triangular structure
of the coefficients, i.e. $a^{(l>m+2n)}_{nm}=0$, which has proved
useful in performing computer symbolic algebraic
manipulations~\cite{LMR02}.

%-----------------------------------
\subsection{Modified CPWE: Closure of the Singlet Bilinear Sector}

Now we compute the contribution of the singlet bilinear sector to
the four-point function by summing the CPWs with the corresponding
fusion coefficients ($\gamma^{uv}_l$) in terms of those of the two
and three-point functions~\footnote{ We can check the consistency
of our conventions by comparing for the energy-momentum tensor
($l=2$). To keep track of the normalization coming from Ward
identities we use $\phi=\frac{1}{\sqrt{2N}}J$ and the canonically
normalized energy-momentum tensor $T=-\frac{1}{2(d-1)S_d}J_2$
~\cite{DO00}, where
$S_d=\frac{2\pi^{\frac{d}{2}}}{\Gamma(\frac{d}{2})}$, to have
$C_{\phi\phi T}=-\frac{\Delta_{\phi}d}{d-1}=-\frac{(d-2)d}{d-1}$.
The fusion coefficient in eq.(4.13) gets multiplied by
$(\frac{1}{\sqrt{2N}})^4$. Then one gets from eq.(4.12) for the
coefficient of the energy-momentum two-point function
$\frac{C_T}{S^2_d}$, with the well known result for the free O(N)
vector model $C_T=\frac{d}{d-1}N$.} as (see~\ref{CPW-fusion}) \beq
(\gamma^{uv}_l)^2\;=\;C^2_{JJJ_l}/C_{J_l}. \eeq Using our previous
results~(\ref{C2} ) and (\ref{C3}) we have \beq \label{gammaUV}
(\gamma^{uv}_l)^2\;=\;16N\,\frac{2^l}{l!}\,\frac{2\,(\delta)^2_l}
{(2\delta-1+l)_l}. \eeq

The result of the direct channel summation ( appendix \ref{Sum})
is the observation that we can expand the first two terms (boxes)
of (\ref{four-free}) in partial waves of the bilinears, in the
s-channel, as (\ref{d11}) \beq \label{double}
16N\,\frac{u^{\delta}}{(r_{12}r_{34})^{2\delta}}\left\{
1+v^{-\delta}\right\} = \frac{1}{(r_{12}r_{34})^{2\delta}}\sum_{l\geq 0,\,
even}(\gamma^{uv}_l)^2
u^{(\Delta_l-l)/2}G^{(l)}(\delta,\delta,\Delta_l;u,v). \eeq

The full connected correlator is obtained then by crossing
symmetry, since the three box diagrams A,B,C transform under
crossing symmetry in the following way,
 \beq
(2 \rightarrow 4, t-channel)(u,v)\rightarrow (v,u) :
(A,B,C)\rightarrow (A,C,B) \eeq \beq (2 \rightarrow 3,
u-channel)(u,v)\rightarrow (1/u,v/u) : (A,B,C)\rightarrow (C,B,A).
\eeq

\begin{figure}
\begin{picture}(300,100)(0,0)
  \put(30,20){\resizebox{130mm}{!}{\includegraphics{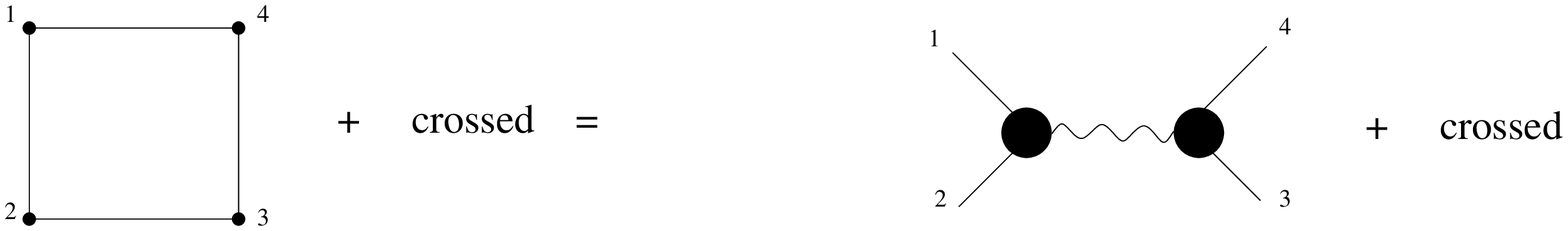}}}
  %\put(220,125){$+\quad crossed$}
  \put(175,42){ $\frac{1}{2}$}
  \put(190,40){\huge $\sum$}
  \put(180,22){\scriptsize $l\geq 0,\,even$}
  \put(220,40){$(\gamma^{uv}_l)^2$}
  \put(285,30){$[J_l]$}
  %\put(320,35){$+\quad crossed$}
\end{picture}
\caption{Free four-point function as a sum partial waves of
minimal twist operators.} \label{f-CPW}
\end{figure}

What we have found amounts to the diagrammatic identity in
fig.~\ref{f-CPW}.

Our rewriting is different from the standard CPWE where the whole
crossing symmetric result is reproduced in each channel. The OPE
of two scalar bilinear J contains the contributions of the
identity, of the conformal blocks of the bilinears (minimal twist)
and also of the ``double-trace" (higher twist) operators starting
with $(\overrightarrow{\varphi}^2)^2$. In the large N analysis,
one can see that when the OPE is inserted in the four-point
function, the identity produces only one piece of the disconnected
part (which goes as $N^2$ ) and the completion comes precisely
form the double-traces (their fusion coefficients squared also
goes as $N^2+O(N)$)~\cite{KP02,AFP00}. It is also easy to see that
the bilinear sector only contributes to the connected part (which
goes as $N$, just like the fusion coefficients squared of this
minimal twist sector, eq.~\ref{gammaUV}).

Now the full disconnected piece is obtained from the Witten graphs
containing two disconnected lines, where the three channels are
included and with no need of additional fields in the bulk of AdS.
One would then expect that for the connected part, something
similar might happen. At leading $1/N$, tree approximation for the
bulk theory, we have a classical field theory where one has to
consider the exchange graphs in each channel separately; trading
bulk exchanges by the corresponding CPW, one should then expect to
write the connected part in terms of only CPWs of the minimal
twist sector in the three channels, with no explicit reference to
contributions from higher-twist/double-trace operators. To our
surprise, this is precisely what we have obtained above!

So that in this way, we rescue the closure of the minimal twist
sector that is in correspondence with a consistent truncation to
the massless sector of the dual HS bulk theory. This result is
valid as well for the bilinear single trace sector of free gauge
theories considered in~\cite{Wit01,SS02,Gop03,Mik02}, and amounts
to a closure of the twist-two sector (without the double-trace
operators this time!) in $d=4$ by including the crossed channels,
in conformity with the expectations for a consistent truncation of
the bulk theory~\cite{SS02,Mik02}.

%As we said before, the CPW are present in the corresponding Witten
%graphs but additional terms arise, there are $log u$ terms that
%should better cancel this time for in the free theory there's no
%room for anomalous dimensions to account for such terms.

%------------------------------------------------------------------------

\section{Degeneracy of the Hologram: IR CFT at d=3}

Now we examine a peculiarity of this O(N) vector model at $d=3$,
which mimics the effect of double-trace deformations of the free gauge
theory ( see e.g.~\cite{GK03} and
references therein).

The canonical dimension of the scalar $J$ is $\Delta=1$. This
value $\Delta_-$ is mapped, via AdS/CFT Correspondence, to a
conformally coupled bulk scalar. However,there is a conjugate
dimension $\Delta_+=2$ which agrees (at leading $1/N$ order)with
the known result for the dimension of $J$ at the interacting IR
critical point. This led  Klebanov and Polyakov to conjecture that
the minimal bosonic HS gauge theory with even spins and symmetry
group $hs(4)$ is related, via standard AdS/CFT methods with the
``conventional" branch $\Delta_+$, to the interacting large N
vector model at its IR critical point. The free theory, UV fixed
point, corresponds then to the other branch $\Delta_-$.

The existence of this IR stable critical point of the O(N) vector
model below four dimension is a well established fact~\footnote{In four
  dimensions the IR fixed point merges with the UV one and the duality is
no longer valid in the way we have just presented. Still one can modify the
  O(N) Vector Model (by gauging) to have a similar holographic scenario~\cite{Sch03}.}.
Standard approaches are the $\epsilon$-expansion in $4-\epsilon$
dimensions which leads to the Wilson-Fisher fixed point and the
large-N expansion which reveals a fixed point at $2<d<4$. Our
analysis will be restricted to the leading $1/N$ results. An
efficient way to perform the large-N expansion is introducing an
auxiliary field $\alpha$ coupled to the vector field via a triple
vertex $\alpha\,\varphi^a\varphi^a$ and then integrate out
$\varphi^a$ which appears now quadratically, to get the effective
action for $\alpha$. The diagrammatic expansion in $1/N$ involves
skeleton graphs with the field $\varphi^a$ running along internal
lines and the triple vertices of two $\varphi$'s with the
auxiliary field~\cite{WK73}.
%-----------------------------------
\subsection{IR Two- and Three-Point Functions}

At leading $1/N$ we keep the free two-point function of the
elementary fields $\varphi^a$, they acquire anomalous dimension of
order $1/N$, and for the auxiliary field $\alpha$ with dimension
$\Delta_+=2$ one can set~\footnote{In what follows and in an abuse
of
 notation, a correlator involving
$\alpha$ is understood to be computed at the IR fixed point, while
the same correlator at UV contains J instead.} \beq \langle
\alpha(x)\alpha(0)\rangle=r^{-2}, \eeq absorbing the
normalizations in the vertex, which becomes~\cite{LR03}(see
appendix~\ref{D'EPP} for notations) \beq
(\frac{z_1}{N})^{1/2}\quad,\quad z_1=-2p(2). \eeq

Analogously, the three-point function form is dictated by the
conformal symmetry \beq \langle \alpha(x_1) \alpha(x_2)
\alpha(x_3)\rangle
= C_{\alpha\alpha\alpha}\;(r_{12}r_{23}r_{31})^{-1} \eeq and \beq
\langle \alpha(x_1) \alpha(x_2)
J_{l\;{\nu_1...\nu_l}}(x_3)\rangle= C_{\alpha\alpha
J_l}\;r^{-2+\delta}_{12}(r_{23}r_{31})^{-\delta}
\;\lambda^{x_3}_{\mu_1...\mu_l}(x_1x_2). \eeq

They are related to the respective free correlators by amputation
relations (Legendre transform). For the two-point function \beq
\langle J(x) J(0)\rangle=-\frac{2N}{p(2)}\langle \alpha(x)
\alpha(0)\rangle^{-1}. \eeq With our choice of normalizations, we
have then to amputate with $\langle \alpha(x)
\alpha(0)\rangle^{-1}$ and multiply by the factor
$(-\frac{2N}{p(2)}) ^{\frac{1}{2}}$ for each scalar leg that is
amputated to go from IR to UV at leading $N$, while the legs
corresponding to the HS current remain the same. To compare this
normalization with that of Klebanov and Witten~\cite{KW99} and
appendix F, we denote their Legendre transformation by
$\Gamma[A]=W[\phi_0]-(2\Delta_+-d)\int A\phi_0$. Then at $d=3$ the
source for $\alpha$ is $\phi_0/\pi$ and that for $J$ is
$A/(2\pi\sqrt{N})$. The amputation is done with the D'EPP formula
and its generalization (see appendix~\ref{D'EPP}). For the
three-point function of the scalars one gets \beq
C_{\alpha\alpha\alpha}=N (-\frac{2}{N p(2)})^{\frac{3}{2}}
v^2(2,\delta,\delta)v(2,1,2\delta-1) \eeq as obtained
in~\cite{LR03,Pet03}. There is a factor $\frac{1} {\Gamma(d-3)}$
that forces the vanishing of the IR three-point function at $d=3$
in correspondence with the vanishing of the bulk coupling in the
HS $AdS_4$ theory~\cite{Pet03,SS03}.

We  extend this amputation procedure to the other three-point
functions to get~\footnote{This time the amputation is done with
the generalization \ref{gdepp} of the D'EPP formula.} \beq
C_{\alpha\alpha
J_l}=2^{l+1}\frac{l!\,(2\delta-1)\,(\delta)_l}{(2\delta-1)_l} \eeq
 which agrees with what was obtained in \cite{LMR04}~\footnote{There is a
  relative factor of 2 due to normalization of the HS current and a missing
  factor $2^l$, by misprint, in
  equation (97) of this paper.} by a different procedure,
 namely computing the four-point function first of the two scalars with two
 elementary fields and then forming the HS current by contracting the two legs
 of the elementary fields acting with derivatives and letting their argument to
 coincide at the end. This computation was
 done at $d=3$, however we corroborate the validity for any $2<d<4$. This
 has the surprising implication that a graph contributing to the four-point
 function above mentioned, which vanishes at $d=3$, does not contribute to the
  HS current correlator at generic $2<d<4$ as well.

%-----------------------------------
\subsection{Scalar Four-Point Function and Fusion Coefficients at IR}

\begin{figure}
\begin{picture}(300,100)(0,0)
  \put(30,20){\resizebox{130mm}{!}{\includegraphics{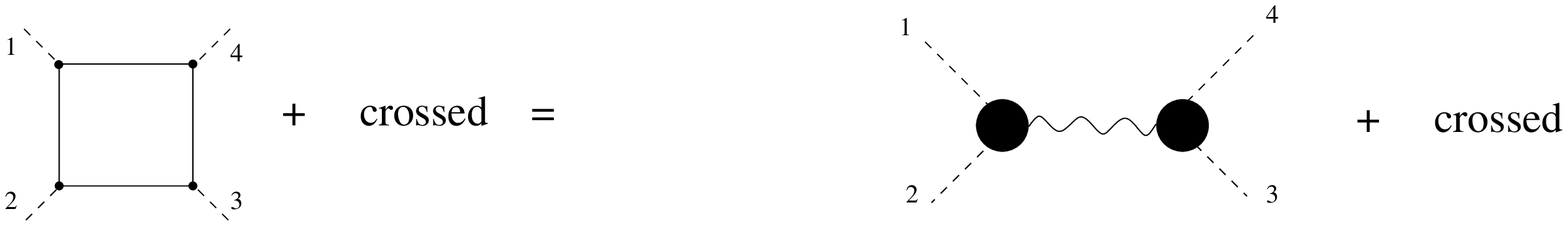}}}
  %\put(220,125){$+\quad crossed$}
  \put(170,42){ $\frac{1}{2}$}
  \put(185,40){\huge $\sum$}
  \put(175,22){\scriptsize $l\geq 2,\,even$}
  \put(215,40){$(\gamma^{ir}_l)^2$}
  \put(280,30){$[J_l]$}
  %\put(320,35){$+\quad crossed$}
\end{picture}
\caption{IR four-point function as a sum of partial waves of
minimal twist operators at $d=3$.} \label{f-CPWIR}
\end{figure}

Let us now examine the implications of the degeneracy of the
hologram for the four-point function at the IR critical point. The
AdS amplitude should involve the same bulk exchange graphs, only
the scalar bulk-to-boundary and and bulk-to-bulk propagators are
switched to the ones with $\Delta_+$. We trade them by CPWs with
the appropriate fusion coefficients that follow from the
amputation program. Therefore we guess the modified CPWE in the
interacting theory as indicated in fig.~\ref{f-CPWIR},

The fusion coefficients \beq
(\gamma^{ir}_l)^2\;=\;C^2_{\alpha\alpha J_l}/C_{J_l}, \eeq using
our previous results from the amputations, are given by \beq
(\gamma^{ir}_l)^2\;=\;\frac{1}{N}\,\frac{2^l}{l!}\,\frac{8(l!)^2\,(\delta)^2_l}
{(2\delta)_{l-1}\,(2\delta)_{2l-1}}. \eeq

That the four-point function at the IR critical point at leading
$1/N$ has precisely this expansion has been shown by
R\"uhl~\cite{Rue04}, by explicit computations at the IR critical
point and the fusion coefficients obtained by extrapolation of
computer algebraic manipulations. What we have analytically found
confirms those results and prove their validity for the whole
range $2<d<4$ where the scalar contribution accounts for the
one-line-reducible graph, both of them being now non-vanishing.
The shadow contribution in the one-line-reducible graph is
canceled by contributions from the box as shown in~\cite{KP02} for
the leading singular term and in~\cite{LMR02} for the full CPW.
The quotient $\gamma^{ir}_l/\gamma^{uv}_l$ for $d=3$ is \beq
\gamma^{ir}_l/\gamma^{uv}_l=l/(2N)~, \eeq which is valid even for
$l=0$ since $\gamma^{ir}_l=0$. Note that in the other
normalization for fields with sources $\phi_0$ and $A$, the ratio
turns out to be equal to $2l$.
%-----------------------------------------------------------------------------
\section{CPWs vs. AdS Exchange Graphs at $d=3$}

The two and three-point functions considered before can be
reproduced from a bulk action, being relevant only up to cubic
terms of the bulk Lagrangian. These have been obtained
in~\cite{LMR04}, assuming a bulk coupling of the HS field with a
bulk current, bilinear in the scalar bulk field and involving up
to l derivatives \footnote{However, what one obtains from the HS
theory for $l=2$ is a bulk energy-momentum involving infinitely
many derivatives. It is still an open issue to see whether both
formulations are equivalent via some field redefinition. This we
believe must first be clarified before trying to explore HS bulk
exchange graphs, the coupling to the scalar is still ambiguous
although their should be fixed by the conformal symmetry.}. In
their scheme there are two different couplings of the HS bulk
field to two bulk scalars, one to reproduce the UV correlators and
another for the IR case, and therefore two different bulk
Lagrangians. We however adopt the view of a unique bulk
Lagrangian, as expected from double trace deformations. It is not
difficult then to realize that the bulk graphs corresponding to
the coupling of the HS fields to the AdS current, bilinear in the
bulk scalar, obtained in ~\cite{LMR04} lead to boundary
three-point functions which are precisely related by amputation of
the scalar legs. This is done in the boundary theory with the
generalized D'EPP formula and in the bulk graph this amounts to
changing the dimension $\Delta_- \leftrightarrow \Delta_+$ of the
bulk-to-boundary propagator of the scalar legs~\cite{Dob98}.

For the four-point function, the CPW expansion obtained is indeed
a step in the ambitious program of bottom to top approach, in
which one uses the knowledge of the boundary CFT to reconstruct
the bulk theory. This is a formidable task, but a Witten graph is
certainly closer to a CPW as we know since the early days of
AdS/CFT, although the precise correspondence has always been
elusive and tricky~\cite{LT98,Liu98}.

Here we will study the scalar exchange and see what happens when
one considers the boundary scalar bilinear to have canonical
dimension $\Delta_-=d-2=1$.

\begin{figure}
\label{fig4}
\begin{picture}(300,100)(0,0)
\put(130,20){\resizebox{70mm}{!}{\includegraphics{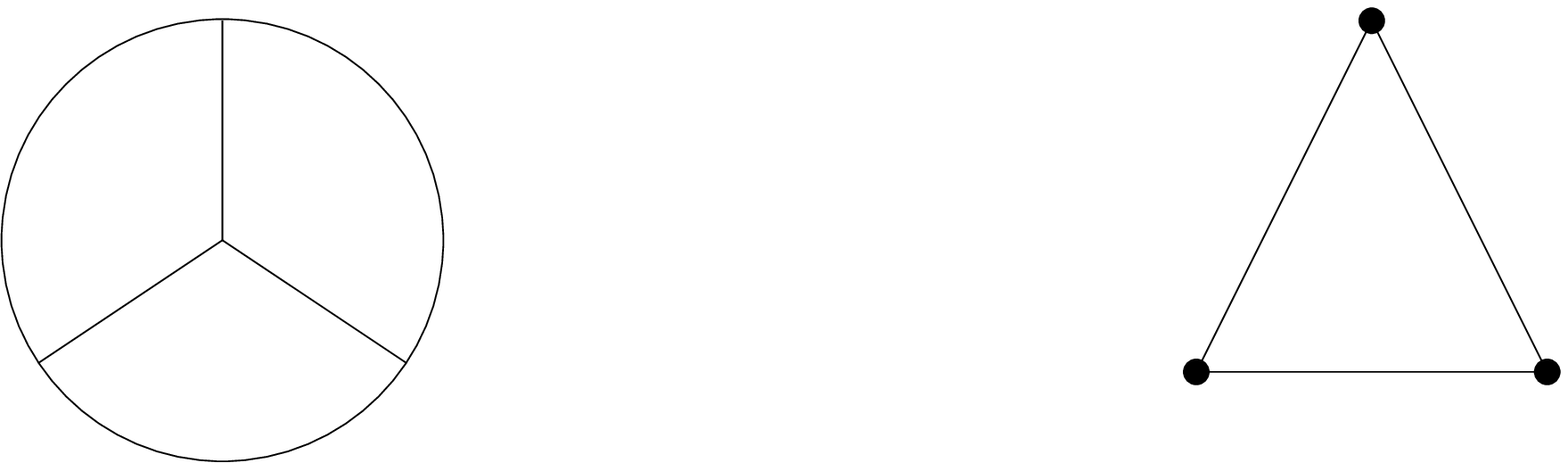}}}
 %\put(60,20){\resizebox{90mm}{!}{\includegraphics{ffig2.eps}}}
  %\put(220,125){$+\quad crossed$}
  \put(155,40){\tiny $g$}
  \put(195,45){$\,=\, g\quad\Gamma (d-3)$}
  \put(155,80){\tiny 1}
  \put(140,60){\tiny $\Delta_-$}
  \put(125,30){\tiny 2}
  \put(140,30){\tiny $\Delta_-$}
  \put(185,30){\tiny 3}
  \put(167,45){\tiny $\Delta_-$}
  \put(302,85){\tiny 1}
  \put(275,20){\tiny 2}
  \put(328,20){\tiny 3}
  %\put(175,22){\scriptsize $l\geq 0,\,even$}
  %\put(215,40){$(\gamma^{IR}_l)^2$}
  %\put(280,30){$[J_l]$}
  %\put(320,35){$+\quad crossed$}
\end{picture}
\caption{Free three-point function and star Witten graph at
$d\rightarrow 3$.} \label{f-star}
\end{figure}

Let us start with the free three-point function. Despite the
success of predicting the vanishing of the scalar three-point
function at the IR critical point and matching with the HS bulk
theory~\cite{Pet03,SS03}, the non-vanishing result for the free
correlator cannot be obtained from a null result via the proposed
Legendre transformation. One is forced to make a regularization,
and the appropriate way turns out to be that the bulk coupling
goes like $g\sim(d-3)$. Here we propose to compute directly with
the canonical dimension and to control the divergence of the
Witten graph by dimensional regularization. The graph is divergent
at $d=3$ (see \ref{three}, \ref{threee}), but a cancellation
against the vanishing coupling gives the correct result for the
free correlator\footnote{Essentially the same cancellation that
occurs for extremal correlators in standard AdS/CFT.}. Starting
with the free correlator and following Gopakumar~\cite{Gop03} in
bringing it to an AdS Witten graph, one gets the identity
sketched~\footnote{In this section the equality sign is to be
understood modulo finite factors that we omit for simplicity. The
precise relation can be read from \ref{three}, \ref{threee}.} in
fig.~\ref{f-star} .
The divergence of the star graph is then controlled by the zero in
$1/\Gamma(d-3)$, rendering the correct result for the
free correlator.

\begin{figure}
\begin{picture}(300,100)(0,0)
  \put(50,20){\resizebox{70mm}{!}{\includegraphics{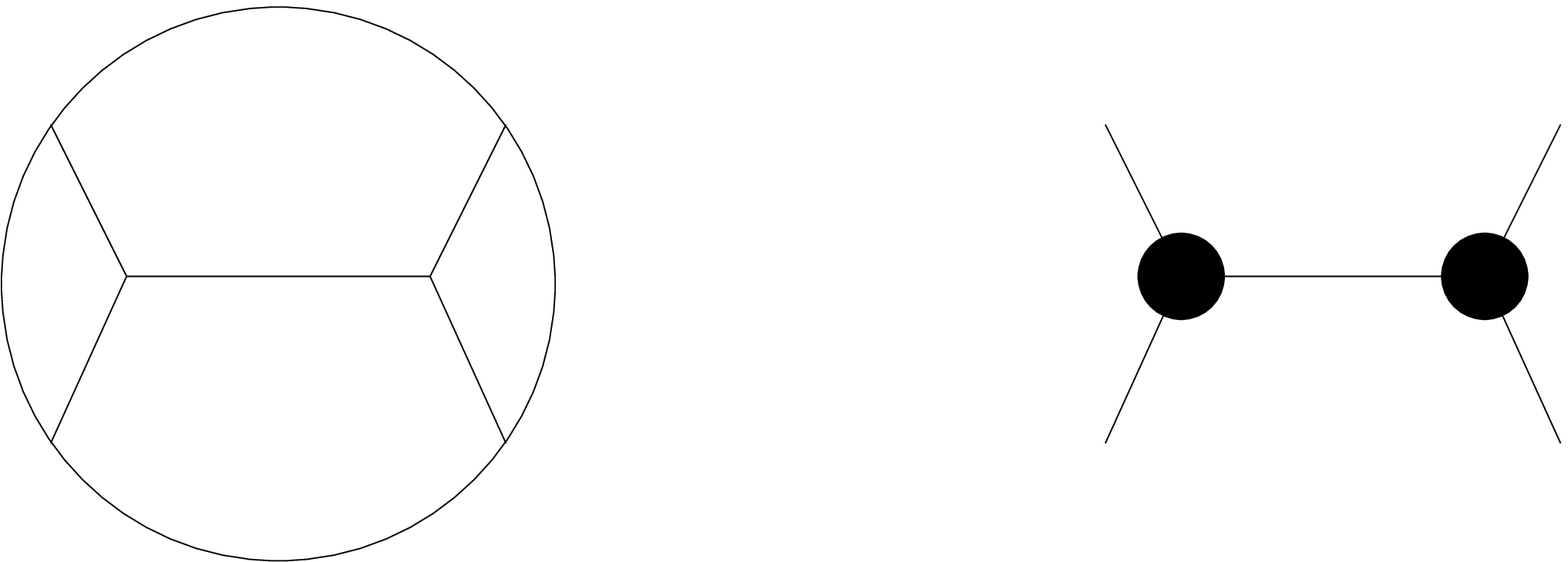}}}
  \put(50,80){\tiny 1}
  \put(62,70){\tiny $\Delta_-$}
  \put(58,55){\tiny $g$}
  \put(107,55){\tiny $g$}
  \put(50,30){\tiny 2}
  \put(62,40){\tiny $\Delta_-$}
  \put(115,80){\tiny 4}
  \put(98,70){\tiny $\Delta_-$}
  \put(115,30){\tiny 3}
  \put(98,40){\tiny $\Delta_-$}
  \put(80,60){\tiny $\Delta_-$}
  \put(150,55){$=$}
  \put(215,45){$[J]$}
  \put(185,80){\tiny 1}
  \put(185,30){\tiny 2}
  \put(250,80){\tiny 4}
  \put(250,30){\tiny 3}
  \put(270,55){$+$\qquad$shadow$}
\end{picture}
\caption{Scalar exchange Witten graph VS. scalar CPWs in the limit
$d\rightarrow 3$} \label{f-scalar}
\end{figure}

Now we  move on and examine the scalar exchange Witten graph, with
the external legs having canonical dimension
$\Delta_-=d-2\rightarrow 1$ and coupling vanishing as $g\sim
(d-3)\rightarrow 0$. A suitable evaluation of this graph, worked
out in ~\cite{HPR000} using Mellin-Barnes representation and
performing a contour integral, is given by a double series
expansion involving three coefficients $a^{(1)}_{nm}$,
$b^{(1)}_{nm}$ and $c^{(1)}_{nm}$ (in \cite{HPR000}, eq.23-26). In
the limit $\Delta=\widetilde{\Delta}=d-2$ and $d\rightarrow 3$,
they all become divergent but only $c^{(1)}_{nm}$ and the first
term in $b^{(1)}_{nm}$ develop a double pole, the rest being less
singular. They cancel against the double zero from $g^2$ and the
final result can be precisely casted into the CPW of the free
scalar $J$ plus its shadow, a scalar of dimension $\Delta_+=2$.
The piece coming from the $c^{(1)}_{nm}$ coefficient goes to the
$c_{nm}(1)$ coefficient of the CPW and the contribution from
$b^{(1)}_{nm}$ produces the shadow term with coefficient
$c_{nm}(2)$ (~\cite{HPR000}, eq.35-36). We end up with a precise
identification in term of CPWs as sketched in fig.~\ref{f-scalar}.

When one continues the crossed channel expansions to get their
contribution in the direct channel one gets $log\,u$ terms but no
non-analytic terms in $(1-v)$~\footnote{This is a simple way to
see that the
  bold identification of the scalar exchange Witten graph with the CPW, as originally
  proposed in~\cite{LT98}, was certainly not correct. In our case we bypass
  this difficulty due to the shadow term, which makes the whole expression
  manifestly ``shadow-symmetric''.}. This
happens both for a Witten graph and for the combined CPW , i.e.
direct plus shadow. The mechanisms that prevent the appearance of
such terms  are different~\cite{HPR00,HPR000}, in one case is due
to some nontrivial hypergeometric identities and in the second
case is due to the presence of the shadow field contribution. In
the above case, both mechanisms coincide and the identification in
terms of CPW is precise (this identification is in general
incomplete, as mentioned before). That is, there is more structure
in the scalar exchange graph than in a generic one and we take
this as a good sign that after all, when computing the infinite
tower of exchange diagrams, many delicate cancellations of
additional terms take place to end up with just the sum of CPWs as
obtained before. In particular, the additional term in the scalar
case is a {\it shadow} contribution which are indeed absent in any
full physical amplitude.

%-----------------------------------------------------------------------------

\section{Conclusion}

We have re-organized the CPWE of the free four-point function of
the scalar singlet bilinear in the natural way one would expect
from AdS/CFT Correspondence; that is, by explicit inclusion of the
crossed channels and involving only CPWs of the minimal twist
sector which is holographically dual to the bulk HS gauge fields.
This result is applicable as well to the corresponding sector of
free gauge theories. Kinematically, double-trace operators are
dual to two-particle bulk states; however, it is hard to see how
such bulk states arise in the tree bulk computation that one has
to perform at leading large-N. We guess that the double-trace
operators arise indirectly, just in the way they show up in the
free O(N) vector model.

In $2<d<4$ dimensions, one can flow (at leading large-N) into the
IR fixed point of the O(N) vector model by just Legendre
transforming. In this way, we have completed the program initiated
in~\cite{Pet03} for the three-point functions. In addition, under
the assumption of a degenerate hologram, i.e. same bulk content
but different asymptotics for the scalar bulk field, the modified CPWE
of the four-point
function was also obtained at IR.

All two- and three-point function coefficients as well as fusion
coefficients were analytically obtained, in some cases
corroborating extrapolations from computer algebraic
manipulations.

For the scalar exchange Witten graph with canonical dimensions, a
funny cancellation occurs and the result can be precisely
identified in terms of CPWs of the corresponding scalar and its
shadow. This reveals more structure than the generic case, and we
hope that such ``accidents" are indeed needed to obtain the full
four-point correlator if one were able to sum the infinite tower
of HS field exchanges.

%-----------------------------------------------------------------------------
\acknowledgments We thank  F.A. Dolan, R. Gopakumar, H. Osborn and
W. R\"uhl for valuable e-mail correspondence. We are also grateful to A.
Torrielli for useful discussions. This work was partially supported
by DFG via GRK 271 and DO 447/3-3.

%-----------------------------------------------------------------------------

\begin{appendix}
 %-------------------------------------------------------
\section{Restrictions from Conformal Invariance}
\label{Conformal}

Conformal invariance dictates the form the three-point function of two
scalars, of  dimension $\Delta_i$, with a totally symmetric
 traceless rank $l$ tensor, of dimension $\Delta$, to be (see e.g.~\cite{FP98})

\[\langle\phi_1(x_1)\phi_2(x_2)O^{(l)}_{\mu_1...\mu_l}(x_3)\rangle\]
\beq \label{three1} =C_{\phi_1\phi_2 O^{(l)}}
\frac{1}{r_{12}^{(\Delta_1+\Delta_2-\Delta+l)/2}r_{13}^{(\Delta
+\Delta_{12}-l)/2}r_{23}^{(\Delta
-\Delta_{12}-l)/2}}\lambda^{x_3}_{\mu_1...\mu_l}(x_1,x_2), \eeq
where \beq
\lambda^{x_3}_{\mu_1...\mu_l}(x_1,x_2)=\lambda^{x_3}_{\mu_1}
(x_1,x_2)...\lambda^{x_3}_{\mu_l}(x_1,x_2)-traces,\quad
\lambda^{x_3}_{\mu}(x_1,x_2)=(\frac{x_{13}}{r_{13}}-\frac{x_{23}}{r_{23}})_\mu,
\eeq and $\Delta_{ij}=\Delta_i-\Delta_j$.

Also the form of the two-point function of the symmetric traceless
rank $l$ tensor, which defines an orthogonality relation with
respect to spin and conformal dimension, is required to be (see e.g.~\cite{FP98})
\beq
\label{two} \langle O^{(l)}_{\mu_1...\mu_l}(x_1)
O^{(l)}_{\nu_1...\nu_l}(x_2)\rangle
= C_{O^{(l)}}\frac{1}{r_{12}^{\Delta}}\,
\;sym\{I_{\mu_1\nu_1}(x)...I_{\mu_l\nu_l}(x)\} \eeq
where
\beq
I_{\mu\nu}(x)=\delta_{\mu\nu}-2\frac{x_{\mu}x_{\nu}}{r} \eeq
is the inversion tensor, related to the Jacobian of the inversion
$x_{\mu}\rightarrow x_{\mu}/r$, and $sym$ means symmetrization and
removal of traces.

The structure of the general four-point conformal correlator is
required to be
 \beq
\langle \phi_1(x_1)\phi_2(x_2)\phi_3(x_3)\phi_4(x_4)\rangle
=\prod^4_{i<j,1}(r_{ij})^{(\Sigma/3-\Delta_i-\Delta_j)/2}F(u,v),\eeq
where $\Sigma=\Delta_1+...+\Delta_4$ and $F$ is an arbitrary
function of the invariant ratios.

 %-------------------------------------------------------
\section{HS Two-Point Function Coefficient}
\label{HS2}

The double sum can be cast into the form
\beq \frac{1}{4}\sum^l_{k=0}(-1)^k{l \choose k}
\frac{(\delta)_{l}\,(\delta)_{2l-k}}{(\delta)_{l-k}}
\sum^l_{s=0}(-1)^s{l \choose s}
\frac{(\delta+k)_s\,(\delta+2l-k)_{-s}}{(\delta)_s\,(\delta+l)_{-s}}.
\eeq
The last sum can be transformed, using elementary identities
such as $(-1)^k{n \choose k}=\frac{(-n)_k}{k!}$ and
$(-z)_n=(-1)^n\frac{1}{(1+z)_{-n}}$, in a terminating generalized
hypergeometric series $_3F_2$ of unit argument
\beq
\sum^l_{s=0}\frac{1}{s!}(-l)_s
\frac{(\delta+k)_s\,(1-\delta-l)_s}{(\delta)_s\,(1-\delta-2l+k)_s}
=\,_3F_2{{-l,\delta+k,1-l-\delta}\choose{\delta,1-2l-\delta-k}}.
\eeq

The evaluation of $_3F_2$ can be done by applying twice the same
identity used in eq.(4.8),
\beq _3F_2{{-n,a,b}\choose{d,e}}=\frac{(e-a)_n}{(e)_n}
\,_3F_2{{-n,a,d-b}\choose{d,1+a-n-e}} \eeq to get
\[
\frac{(k-l)_l}{(1-2l-\delta+k)_l}\frac{(1-k+l)_k}{(1-k)_k}
\,_3F_2{{-k,-l,2\delta+l-1}\choose{\delta,-l}}
\]
\beq = \frac{(k-l)_l}{(1-2l-\delta+k)_l}\frac{(1-k+l)_k}{(1-k)_k}
\,\frac{(1-\delta-l)_k}{\delta_k}=(-1)^k\frac{l!}{(\delta)_k}
{(\delta+l)_{l-k}} \eeq
where the $_3F_2$ reduced to an ordinary
$_2F_1$ of unit argument evaluated with the Chu-Vandermonde
formula~\cite{AAR}.

The sum that remains to be done reduces then again to a
terminating ordinary hypergeometric of unit argument that is
evaluated as before
\[
\frac{1}{4}\,l!\,(\delta)_l\sum^l_{k=0}{l \choose k}
\frac{1}{(\delta)_k\,(\delta+l)_{-k}}=\frac{1}{4}\,l!\,(\delta)_l\;
_2F_1{{-l,1-\delta-l}\choose{\delta}}\]
\beq =\frac{1}{4}\,l!\,(2\delta-1+l)_l. \eeq

%-------------------------------------------------------
\section{CPW Recurrences}
\label{CPW-rec}

Inserting the OPE~\footnote{The OPE involves the sum over the
complete set of quasi-primaries. We consider no degeneracies for
simplicity, ie. no additional labels apart from ($\Delta,l)$.}
(\ref{OPE}) and using the orthogonality relation (\ref{two}) we
have for the action of the derivative operator

\[\langle\phi_1(x_1)\phi_2(x_2)\;O^{(l)}_{\mu_1...\mu_l}(x_3)\rangle\]
\beq \label{der} =\frac{C_{\phi_1\phi_2 O^{(l)}}}{C_{O^{(l)}}}
\frac{1}{r_{12}^{(\Delta_1+\Delta_2-\Delta+l)/2}}
C^{(l)}(x_{12},\partial_{x_2})_{\nu_1...\nu_l} \langle
O^{(l)}_{\nu_1...\nu_l}(x_2)\;
O^{(l)}_{\mu_1...\mu_l}(x_3)\rangle. \eeq

Inserting now the OPE (\ref{OPE}) in the scalar four-point
function one gets for the contribution of $O^{(l)}$ and its
descendants

\bea \langle \phi_1(x_1)\phi_2(x_2)\phi_3(x_3)\phi_4(x_4)\rangle
\sim \frac{C_{\phi_1\phi_2
O^{(l)}}}{C_{O^{(l)}}}\frac{1}{r_{12}^{(\Delta_1+\Delta_2-\Delta+l)/2}}\\
\times C^{(l)}(x_{12},\partial_{x_2})_{\mu_1...\mu_l} \langle
O^{(l)}_{\mu_1...\mu_l}(x_2)\;\phi_3(x_3)\phi_4(x_4)\rangle. \eea

In order to be able to act as before with the derivative operator
on a two-point function, one has to re-write the $x_2$-dependence
in $\langle
O^{(l)}_{\mu_1...\mu_l}(x_2)\;\phi_3(x_3)\phi_4(x_4)\rangle$ in a
suitable way. This is achieved by introducing the ``shadow''
operator (conformal partner) $O^{* (l)}$, a ``conventional''
operator with labels $(\Delta^*,l)=(d-\Delta,l)$

\beq O^{(l)}_{\mu_1...\mu_l}(x_2)= \int d^dx \;\langle
O^{(l)}_{\mu_1...\mu_l}(x_2)\;O^{(l)}_{\nu_1...\nu_l}(x)\rangle \;
O^{*(l)}_{\nu_1...\nu_l}(x). \eeq

Inserting this relation and using (\ref{der}) one gets \beq
\langle \phi_1(x_1)\phi_2(x_2)\phi_3(x_3)\phi_4(x_4)\rangle \sim
\int d^dx \; \langle
\phi_1(x_1)\phi_2(x_2)\;O^{(l)}_{\mu_1...\mu_l}(x)\rangle
\;\langle
O^{*(l)}_{\mu_1...\mu_l}(x)\;\phi_3(x_3)\phi_4(x_4)\rangle. \eeq

In all, one has just inserted the projection
operator~\cite{KP02,LT98}
\beq \mathcal{P}_{_l}\;=\;\int d^dx
\; O^{(l)}_{\mu_1...\mu_l}(x)|0\rangle \;\langle
0|O^{*(l)}_{\mu_1...\mu_l}(x). \eeq

 The integrand can
be cast into a form involving Gegenbauer polynomials after
contraction of Lorentz indices, and using their recurrence
relations one gets the following recurrences\footnote{In fact, by
the previous procedure one gets in addition the contribution from
the shadow operator. What follows is valid for the direct
contribution.}~\cite{DO00}
\[G^{(l)}(b,e,S;u,v)\]
\[=\frac{1}{2}\frac{S+l-1}{d-S+l-2}\left\{\frac{d/2-e-1}{f+l-1}\left(
vG^{(l-1)}(b+1,e+1,S;u,v)-G^{(l-1)}(b,e+1,S;u,v)\right)\right.\]
\[\left.+\frac{d/2-f-1}{e+l-1}\left(
G^{(l-1)}(b,e,S;u,v)-G^{(l-1)}(b+1,e,S;u,v)\right)\right\}\]
\[-\frac{1}{4}\frac{(S+l-1)(S+l-2)}{(d-S+l-2)(d-S+l-3)}\frac{(d/2-e-1)
(d/2-f-1)}{(f+l-1)(e+l-1)}\frac{(l-1)(d+l-4)}{(d/2+l-2)(d/2+l-3)}\]
\beq \label{recurrences}  uG^{(l-2)}(b+1,e+1,S;u,v), \eeq
with $S=e+f+l$. The starting point is the scalar result that in the direct
channel limit, $u, 1-v \sim 0$ is given by the double power expansion
\beq \label{scalar}
G^{(0)}(b,e,S;u,v)=\sum^{\infty}_{m,n=0}\frac{(S-b)_n(S-e)_n}{(S+1-d/2)_n}
\frac{(b)_{n+m}(e)_{n+m}}{(S)_{2n+m}}\frac{u^n}{n!}\frac{(1-v)^m}{m!}.
\eeq

%-------------------------------------------------------
\section{UV Fusion Coefficients}
\label{Sum}

Using the double expansion in the direct channel limit  $u, 1-v
\sim 0$, the sum we have to perform on the RHS of (\ref{double})
is \beq \sum_{l\geq 0,\,even}(\gamma^{uv}_l)^2\;a^{(l)}_{nm} \eeq
with $a^{(l)}_{nm}$ and $(\gamma^{uv}_l)^2$ given in
(\ref{cpwcoeff}) and (\ref{gammaUV}), respectively.

First one can perform the sum over $l$, due to the triangular
structure the sum is up to $m+2n$, and sum over all $l's$ writing
\beq
(\gamma^{uv}_l)^2=[1+(-1)^l]\frac{(\delta)_l^{\;\;2}}{(2\delta+l-1)_l}
\frac{16N}{a^{(l)}_{0l}}. \eeq

Now, after straightforward manipulations\footnote{A term
$(2\delta+2l)_{-1}$ involving $2l$ must be casted into
$\frac{1}{2}(\delta+l+\frac{1}{2})_{-1}$ .}, the sum can be casted
in terms of a terminating well-poised generalized hypergeometric
$_3F_2$ of argument $\pm 1$ as
\[\sum_{l\geq 0,\,even}(\gamma^{uv}_l)^2\;a^{(l)}_{nm}= 16N\,
(\delta)_{m+n}\sum^{n}_{s=0}(-1)^s{n \choose s}
 \frac{(\delta)_{m+n+s}}{(2\delta)_{m+n+s}}\,\]
\beq
 \times\,\left\{
 _3F^-_2{{-m-n-s,\delta+\frac{1}{2},2\delta-1}
  \choose{2\delta+m+n+s,\delta-\frac{1}{2}}}
  +\,_3F^+_2{{-m-n-s,\delta+\frac{1}{2},2\delta-1}
  \choose{2\delta+m+n+s,\delta-\frac{1}{2}}}
  \right\}.
\eeq

The evaluation at $-1$, with the particular case of a corollary of Dougall's
formula (\cite{AAR}, pp.148)
\beq _3F^-_2{{a,1+\frac{a}{2},b}
\choose{\frac{a}{2},1+a-b}}=\frac{(1+a)_{-b}}{(\frac{1}{2}+
\frac{a}{2})_{-b}}\eeq 
gives \beq
\frac{(2\delta)_{m+n+s}}{(\delta)_{m+n+s}}\,, \eeq so that the first
part is \beq 16N\,(\delta)_{m+n}\sum^{n}_{s=0}(-1)^s{n \choose s}
=16N\, (\delta)_{m+n} \,\delta_{n,0}=16N\,(\delta)_{m}
\,\delta_{n,0}\,. \eeq

The evaluation at $+1$ done with Dixon's identity (\cite{AAR}, pp.72),
which can also be derived from Dougall's formula,
\beq
 _3F^+_2{{a,b,c}
\choose{1+a-b,1+a-c}}=\frac{(1+a)_{-b}\,(1+a)_{-c}\,
(1+\frac{a}{2})_{-b-c}}{(1+\frac{a}{2})_{-b}\,(1+\frac{a}
{2})_{-c}\,(1+a)_{-b-c}} \eeq
produces a factor $(0)_{m+n+s}$
which vanishes for $m+n+s\neq 0$. At $m=n=0$ (this forces $s=0$) one
gets $1$, so that the second part contributes
\beq
16N\,(\delta)_{m+n}\,\delta_{n,0}\,\delta_{m,0}\,=\,16N\,
\delta_{n,0}\,\delta_{m,0}. \eeq

Finally, we find the equality \beq \label{d11}
16N\,\delta_{n,0}\left\{\delta_{m,0}+(\delta)_m\right\}\,=\,\sum_{l\geq
0,\,even} (\gamma^{uv}_l)^2\;a^{(l)}_{nm},\eeq

which corresponds to the double expansion of equation
\ref{double}.

%-------------------------------------------------------
\section{D'EPP formula and star Witten graph}
\label{D'EPP}

The inverse kernels are defined according to
\beq p(\lambda)\int
d^dx_3\;
r^{-\lambda}_{13}\;r^{-d+\lambda}_{23}\,=\,\delta^d(x_{12}), \eeq
where \beq p(\lambda)=p(d-\lambda)=\pi^{-d}\frac{\Gamma(\lambda)
\,\Gamma(d-\lambda)}{\Gamma(\frac{d}{2}-
\lambda)\,\Gamma(\lambda-\frac{d}{2})}. \eeq

The D'Eramo-Parisi-Peliti formula~\cite{Pet03,FP98,D'EPP71} reads
\beq \label{depp} \int d^dx_4\;
r^{-\delta_1}_{14}\;r^{-\delta_2}_{24}\;r^{-\delta_3}_{34}\,=\,
v(\delta_1,\delta_2,\delta_3)\;r^{-\frac{d}{2}+\delta_3}_{12}
\;r^{-\frac{d}{2}+\delta_1}_{23}\;r^{-\frac{d}{2}+\delta_2}_{13}\;,
\eeq
where $\delta_1+\delta_2+\delta_3=d$ and
\beq
v(\delta_1,\delta_2,\delta_3)\,=\,\pi^{\frac{d}{2}}\frac{\Gamma
(\frac{d}{2}-\delta_1)\,\Gamma(\frac{d}{2}-\delta_2)\,\Gamma
(\frac{d}{2}-\delta_3)}{\Gamma(\delta_1)\,\Gamma(\delta_2)
\,\Gamma(\delta_3)}. \eeq

We also need a generalization of D'EPP~\cite{FP98}, obtained by
differentiation,
\[\int d^dx_4\;
r^{-\delta_1}_{14}\;r^{-\delta_2}_{24}\;r^{-\delta_3}_{34}\,
\lambda^{x_1}_{\mu_1...\mu_s}(x_4,x_2)=\]
\beq \label{gdepp}
\frac{(\frac{d}{2}-\delta_2)_s}{(\delta_1)_s}\,
v(\delta_1,\delta_2,\delta_3)\;r^{-\frac{d}{2}+\delta_3}_{12}\;
r^{-\frac{d}{2}+\delta_1}_{23}\;r^{-\frac{d}{2}+\delta_2}_{13}\,
\lambda^{x_1}_{\mu_1.\mu_s}(x_3,x_2)\;. \eeq

The star Witten graph with scalar legs of generic dimensions
$\Delta_i$ ($i=1,2,3$) is given by (see e.g. \cite{D'HF02})

\beq \label{three} \frac{a(\Delta_1,\Delta_2,\Delta_3)}
{r_{12}^{(\Delta_1+\Delta_2-\Delta_3)/2}r_{13}^{(\Delta_1
+\Delta_3-\Delta_2)/2}r_{23}^{(\Delta_2 +\Delta_3-\Delta_1)/2}},
\eeq

where

\beq \label{threee}a(\Delta_1,\Delta_2,\Delta_3)=
\frac{1}{2\pi^d}\Gamma(\frac{\Delta_1+\Delta_2+\Delta_3-d}{2})
\frac{\Gamma(\frac{\Delta_1+\Delta_2-\Delta_3}{2})
\Gamma(\frac{\Delta_1+\Delta_3-\Delta_2}{2})
\Gamma(\frac{\Delta_2+\Delta_3+\Delta_1}{2})}{\Gamma(\Delta_1-\frac{d}{2})
\Gamma(\Delta_2-\frac{d}{2})\Gamma(\Delta_3-\frac{d}{2})}. \eeq

%-------------------------------------------------------
\section{Regularized kernels}
\label{DeltapmP}

The aim of this part is to fix notation and thereby to summarize
the facts concerning the reconstruction of the bulk fields out of
its two types of asymptotics along the line of \cite{KW99,mueck}.
Our presentation contains some new elements, insofar as we
exclusively rely on convergent position space integrals. From them
we will be able to $derive$ the analytic continuation rules which
usually appear a posteriori to give meaning to naively divergent
integrals.

In $AdS_{d+1}$ the scalar on shell bulk field $\phi (x)$, with
$x=(z,\vec x),~z\geq 0$ denoting
Poincar\'e coordinates, has the near boundary asymptotics
\footnote{In the following we assume a suitable rapid falloff of $A$ and $\phi
  _0$ for $\vert\vec x\vert\rightarrow\infty $.}, see e.g. \cite{KW99,malda}
\beq
\phi (x)~=~z^{\Delta _-}(\phi _0(\vec x)+{\cal O}(z^2))~+~z^{\Delta _+}(A(\vec
x)+{\cal O}(z^2))~,\label{phias}
\eeq
where $\Delta _{\pm}=\frac{d}{2}\pm\sqrt{\frac{d^2}{4}+m^2}$.
The standard bulk to bulk propagators obey ($\Delta =\Delta_{\pm}$)
\bea
(\Box _x -m^2)G_{\Delta}(x,x')&=&-
    g^{-\frac{1}{2}}~\delta(x,x')~,\nonumber\\
G_{\Delta}(x,x')&=&z'^{\Delta }G^0_{\Delta}(x,\vec{x'})+{\cal
  O}(z'^{\Delta +2})\label{prop}~.
\eea
Using (\ref{phias}),(\ref{prop}) and Gauss theorem one gets with
fixed $z'>0$ \cite{giddings,sieg,dss}
\bea
\phi (x)&=~\int d^d\vec{x'}&\{(\Delta-\Delta
_-)\phi_0(\vec{x'})~G^0_{\Delta}(x,\vec{x'})~z'^{\Delta_-+\Delta -d}+{\cal
O}(z'^{\Delta_-+\Delta -d+2})\nonumber\\
&&+(\Delta-\Delta
_+)A(\vec{x'})~G^0_{\Delta}(x,\vec{x'})~z'^{\Delta_++\Delta -d}+{\cal
O}(z'^{\Delta_++\Delta -d+2})\}~.\label{boundrep}
\eea
Since always $\Delta_+\geq \frac{d}{2}$, for the choice $\Delta =\Delta_+$
both ${\cal O}$-terms go to zero
for $z'\rightarrow 0$. Choosing instead $\Delta =\Delta_-$, the vanishing of
both ${\cal
  O}$-terms  requires $\Delta _->\frac{d-2}{2}$, i.e. just the unitarity
bound. Altogether for $\frac{d-2}{2}<\Delta _-<\frac{d}{2}<\Delta _+$  one gets
\beq
\phi (x)~=~\int
d^d\vec{x'}~\phi_0(\vec{x'})~K_{\Delta_+}(x,\vec{x'})~=~\int
d^d\vec{x'}~A(\vec{x'})~K_{\Delta_-}(x,\vec{x'})~,\label{boundrep2}
\eeq
with
\beq
K_{\Delta_{\pm}}(x,\vec{x'})~=~(2\Delta_{\pm}-d)\lim _{z'\rightarrow 0}
  z'^{-\Delta_{\pm}}~G_{\Delta_{\pm}}(x,x')~=~\frac{\Gamma (\Delta
    _{\pm})}{\pi^{\frac{d}{2}}\Gamma (\Delta_{\pm} -\frac{d}{2})}~\frac{z^{\Delta
      _{\pm}}}{(z^2+(\vec x-\vec{x'})^2)^{\Delta_{\pm}}}~.\label{bulkbound}
\eeq
The reconstruction of the asymptotics (\ref{phias})  from the first eq. in
(\ref{boundrep2})  is given by
\bea
\int
d^d\vec{x'}~\phi_0(\vec{x'})~K_{\Delta_+}(x,\vec{x'})~=~z^{\Delta_-}~\phi_0(\vec
x)~(1+{\cal O}(z^2)+\dots + {\cal O}(z^{2k}))~~~~~~~~~~~~~~~~~~~~~\\
+~\frac{\Gamma (\Delta _+)}{\pi ^{\frac{d}{2}}\Gamma (\Delta
  _+-\frac{d}{2})}~~z^{\Delta _+}~\left (\int
  d^d\vec{x'}~\frac{\phi_0(\vec{x'})-\phi_0(\vec{x})-\dots
    -\frac{((\vec{x'}-\vec
      x)\vec{\partial})^{2k}}{(2k)!}\phi_0(\vec{x})}{\vert \vec{x'}-\vec
    {x}\vert ^{2\Delta_+}}~+~{\cal O}(z^{2})\right ),\nonumber\label{+as}
\eea
where $k$  is the largest integer smaller than $\Delta _+-\frac{d}{2}$.
Similarly one finds from the second representation of $\phi (x)$ in
(\ref{boundrep2}) for $\frac{d-2}{2}<\Delta _-<\frac{d}{2}$
\bea
\int
d^d\vec{x'}~A(\vec{x'})~K_{\Delta_-}(x,\vec{x'})~=~z^{\Delta_+}~A(\vec
x)~(1+{\cal O}(z^{2(\Delta _--\frac{d}{2}+1)}))~~~~~~~~~~~~~~~~~~~~~~~~~~~~~\\
+~\frac{\Gamma (\Delta _-)}{\pi ^{\frac{d}{2}}\Gamma (\Delta
  _--\frac{d}{2})}~~z^{\Delta _-}~\left (\int
  d^d\vec{x'}~\frac{A(\vec{x'})}{\vert \vec{x'}-\vec
    {x}\vert ^{2\Delta_-}}~+~{\cal O}(z^{2})\right ).\nonumber\label{-as}
\eea
We are mainly interested in the situation where $both$ $\Delta _{\pm}$ are
above the unitarity bound, then $k=0$ and $A$ and $\phi _0$ are related via the
convergent position space integrals
\beq
A(\vec x)=\frac{\pi ^{-\frac{d}{2}}\Gamma (\Delta_+)}{\Gamma (\Delta
  _+-\frac{d}{2})}\int d^d\vec {x'}\frac{\phi_0(\vec{x'})-\phi_0(\vec
  x)}{\vert \vec{x'}-\vec x\vert ^{2\Delta _+}}~,~~~\phi_0(\vec x)\frac{\pi ^{-\frac{d}{2}}\Gamma (\Delta_-)}{\Gamma (\Delta
  _--\frac{d}{2})}\int d^d\vec {x'}\frac{A(\vec{x'})}{\vert
\vec{x'}-\vec x\vert ^{2\Delta _-}}~.\label{aphi}
\eeq
Comparing the  first formula in (\ref{aphi}), containing a subtraction, with
the analytic continuation from $\Delta <\frac{d}{2}$ of the corresponding
formula without subtraction, we find for $\frac{d}{2}<\Delta <\frac{d}{2}+1$
\beq
\int d^d\vec {x'}\frac{\phi_0(\vec{x'})-\phi_0(\vec
  x)}{\vert \vec{x'}-\vec x\vert ^{2\Delta }}~=~\left (
\int d^d\vec {x'}\frac{\phi_0(\vec{x'})}{\vert \vec{x'}-\vec x\vert ^{2\Delta
  }}\right )_{\makebox{\scriptsize continued}}~.\label{cont}
\eeq To check (\ref{cont}) one has to split the integral in two
parts $\vert \vec{x'}-\vec x\vert < K$ or $>K$, use the falloff
property of $\phi _0$ at $\vert \vec{x'}\vert \rightarrow\infty$
and to send the arbitrary auxiliary scale $K$ to infinity $after$
the continuation. Remarkably, the singularity of the r.h.s. for
$\Delta \rightarrow \frac{d}{2}-0$ due to the short distance
behavior is reproduced on the l.h.s. for $\Delta \rightarrow
\frac{d}{2}+0$  via the infrared behavior of the subtraction term.
\end{appendix}

%----------------------------------------------------------------------------

\end{document}